\documentclass[aps,twocolumn,final,balancelastpage,superbib,floatfix,byrevtex,showpacs]{revtex4}
\usepackage{graphics}
\usepackage{amsmath}
\usepackage{amssymb}
\usepackage{amsfonts}
\usepackage{graphicx}

\def\eg{{\em e.g.}}

\def\etc{{\em etc.}}

\def\sizeFIG{.75\linewidth}
\def\full{\protect\mbox{------}}

\def\vare{\varepsilon^{\sf e}}
\def\varm{\varepsilon^{\sf m}}
\def\vari{\varepsilon^{\sf i}}

\begin{document}

\title{A formula for dielectric mixtures}
\author{Enis Tuncer}
\email{enis.tuncer@physics.org}
\noaffiliation
\begin{abstract}
Dielectric properties of material mixtures are of importance in diagnostics, characterization and design of systems in various engineering fields. In this Letter, we propose a peculiar dielectric mixture expression, which is based on the dielectric relaxation phenomena and the spectral density representation [E. Tuncer  {\em J. Phys. Condens. Matter} {\bf 17}(12) L125 (2005)]. The expression is tested on several composite systems. Results illustrate that the proposed expression can be used to obtain valuable structural informations in composites, even for highly filled, bi-percolating, systems. Lastly, the proposed expression is an alternative to other existing homogenization formulas in the literature.
\end{abstract}
\maketitle

Predicting the dielectric properties of mixtures with or without knowing the composition of the mixture is an old problem in science and technology~\cite{SihvolaBook}. The structure-property relationship in composite materials has been sought by researchers in various fields \eg, physics, electrical engineering, materials science, geophysics, pharmaceutics, biology, \etc, in order to comprehend the nature of material mixtures for many decades. Such a deep understanding would be of great value to be able to calculate either the dielectric constant of a mixture of substances of known dielectric constants or, knowing the dielectric constants of a mixture of two components and that of one of the components, to calculate the dielectric constant of the other~\cite{Lowry1927}, or even knowing the dielectric constants of a mixture of two components and that of two components to estimate the morphology of the mixture~\cite{TuncerSpectralPRB}. Recently, it has been shown~\cite{Tuncer2005JPCMLET} that the application of the knowledge in dielectric relaxation phenomena and the theory developed by Bergman~\cite{Bergman4} and Milton~\cite{Milton1981}, significant information could be obtained from the frequency dependent dielectric properties of a mixture when the dielectric properties of he constituents were known. In this Letter, we present a peculiar dielectric mixture formula, and utilize it to analyze the dielectric data of oil-in-water emulsion (O/W)~\cite{Smith1990}, brine-porous rock (B/PR)~\cite{Kenyon} and glass-beads-paraffin wax (GB/PW)~\cite{YoungsNano} systems.

\begin{figure}[t]
  \centering
  \includegraphics[width=\sizeFIG]{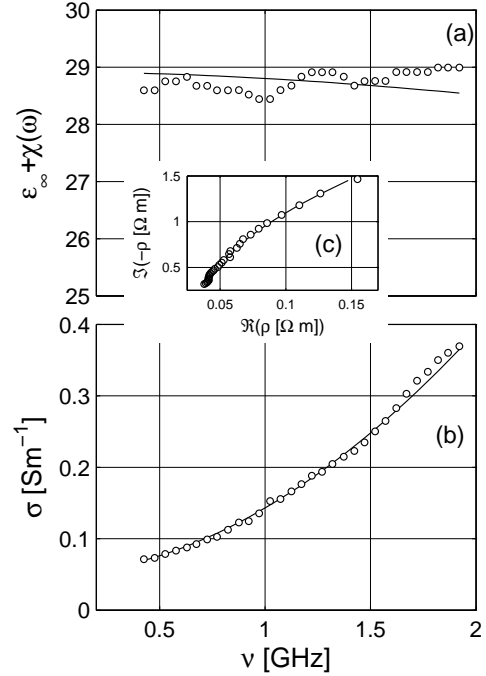}
  \caption{Frequency dependent (a) relative dielectric permittivity and (b) conductivity $\sigma$ of the oil-in-water emulsion system. The Argand plot of the complex resistivity $\rho$ is shown in (c). The measured and modeled data are presented with the open-symbols ($\circ$) and the solid line (\full), respectively.}
  \label{fig:fig1}
\end{figure}

We first introduce a `scaled' dielectric permittivity $\xi$ of the two-component mixture system as,
\begin{eqnarray}
  \label{eq:myequation1}
  \xi&=&(\vare-\varm)(\vari-\varm)^{-1}
\end{eqnarray}
where the subscripts `{\sf e}', `{\sf m}' and `{\sf i}' denote the complex dielectric permittivities $\varepsilon$ with ohmic loss contributions of the effective, matrix and inclusion medium, respectively. Observe that $\xi$ varies between 0 and 1. Second, the complex permittivities are expressed as,
\begin{eqnarray}
  \label{eq:1}
  \varepsilon^{\sf k}=\varepsilon^{\sf k}_\infty+\chi^{\sf k}(\omega)+\sigma^{\sf k}(\imath\omega\varepsilon_0) \quad \text{where} \quad {\sf k}=\{{\sf e,m,i}\}   
\end{eqnarray}
Here $\varepsilon_\infty$ and $\sigma$ are the dielectric permittivity at optical frequencies and the ohmic conductivity of the material. The parameter $\chi$ is the dielectric polarization term, called the susceptibility, which is a complex and frequency $\nu$ dependent quantity, $\omega$ in Eq.~\eqref{eq:1} is the angular frequency ($\omega=2\pi\nu$), and $\imath=\sqrt{-1}$.  We propose that the dielectric permittivity of a mixture can in principle be written as the dielectric relaxation with the `scaled' permittivity replacing the dielectric permittivity of the material,\cite{Tuncer2005JPCMLET}
\begin{eqnarray}
  \label{eq:2}
  \xi=\xi_s+\Delta\xi[1+(\varpi x)^\alpha]^{-\beta}
\end{eqnarray}
Here $\xi_s$ is the fraction of inclusion phase `{\sf i}' forming a percolating network in the direction of the applied field---indicating the fraction of infinite cluster(s), which does not contribute to the interfacial polarization. $\Delta\xi$ is the fraction of inclusion phase that is related to the isolated, deserted, clusters, which contribute to the polarization of the system. The permittivities of the constituents construct the so-called probing spectral frequency $\varpi(\equiv\vari/\varm-1)$. The quantity $x$ is the spectral parameter, contains information on the effective depolarization factors inclusions. The parameters $\alpha$ and $\beta$  depend on the geometrical description of the system. Consequently the parameters $x$, $\alpha$ and $\beta$ yield the distribution of depolarization factors for the given system in hand\cite{HNdist}.  The concentration $q$ of the inclusions are simply defined with $q=\xi_s+\Delta\xi$; as a result for non-percolating systems $q\equiv\Delta\xi$ and completely percolating systems $q\equiv\xi_s$. The defined parameters in Eq. \eqref{eq:2} are positive and smaller than one, $0<\{\xi_s,\,\Delta\xi,\,x,\,\alpha,\,\beta\}\le1$. Rearranging Eq.~(\ref{eq:2}), we propose a dielectric mixing law as follows
\begin{eqnarray}
  \label{eq:myequation2}
  \vare=\varm+(\vari-\varm)
  \{\xi_s+\Delta[1+(\varpi x)^\alpha]^{-\beta}\}
\end{eqnarray}
\begin{table}[t]
  \caption{The non-linear complex least-squares fit results.}
\label{table}
  \centering
  \begin{tabular*}{\linewidth}{l@{\extracolsep{\fill}}rrrrr}
    \hline
    System & $\xi_s$ & $\Delta\xi$ & $x$ & $\alpha$ & $\beta$ \\
    \hline
    O/W   & 0.650 & $3\times10^{-6}$ & $4\times10^{-8}$& 0.962 & 0.768\\ 
    B/PR  & 0.064 & 0.123 & 0.032 & \multicolumn{1}{l}{1} & 0.741\\
    GB/PW & 0.025 & 0.269 & 0.152 & \multicolumn{1}{l}{1} & \multicolumn{1}{l}{1}\\
    \hline
    \hline
  \end{tabular*}
\end{table}

We have selected three different systems to test the proposed expression in Eq.~\eqref{eq:myequation2}; ({\em i}) oil-in-water~\cite{Smith1990}, ({\em ii}) brine-in-porous-rock~\cite{Kenyon} and ({\em iii}) metal-coated glass-beads-in-paraffin wax \cite{YoungsNano} composite systems. The data for the first two systems are digitized from the appropriate references. In the analysis we present the complex resistivity $\rho[\equiv(\imath\varepsilon_0\varepsilon\omega)^{-1}]$ level \cite{RossBJP} of the systems as well to instruct the significance of this representation in lossy dielectric composites.


\citet{Smith1990} studied the dielectric properties of emulsion systems and applied the Bruggeman unsymmetric mixture formula~\cite{Bruggeman1935} to investigate the behavior of their system, in which the oil was assumed to be the inclusion phase with lossless permittivity, and the frequency dependent complex permittivity of the water (saline solution) was calculated as presented in Ref. \onlinecite{Smith1990}. The permittivity of the composite system and  response estimated with Eq.~\eqref{eq:myequation2} by employing a non-linear complex least squares algorithm are shown in Fig.~\ref{fig:fig1}. The fit parameters are listed in Table~\ref{table}. The large discrepancy between measured and modeled data in Fig.~\ref{fig:fig1}a is due to the digitization procedure, however, the model values are within the range of the measured permittivity. The complex conductivity $\sigma$ estimate on the other is in accord with the measurement and better than Bruggeman's expression\cite{Bruggeman1935} as shown in Ref. \onlinecite{Smith1990}. Since the system shows conductive behavior, the Argand plot of the complex resistivity is also presented in Fig.~\ref{fig:fig1} to indicate the merit of the fitting expression. The concentration $q(\approx0.66)$ estimated indicates that the saline solution content is incompatible with the value given by  \citet{Smith1990}, $q=0.5$. We therefore argue that although the mixture was composed of 1:1 oil and water, not all saline solution was contributing to the conductivity as suggested by \citet{Smith1990}. This observation means that the emulsifier contant was higher than desired in the prepared mixture.

\begin{figure}[t]
  \centering
  \includegraphics[width=\sizeFIG]{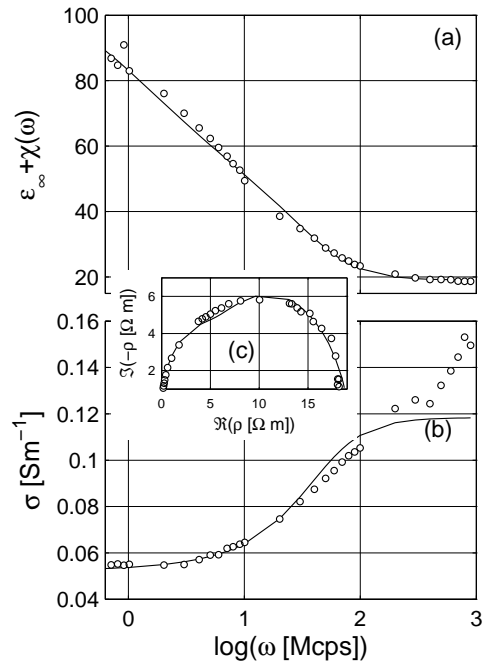}
  \caption{Frequency dependent (a) relative dielectric permittivity and (b) conductivity $\sigma$ of the brine-porous-rock system. The Argand plot of the complex resistivity $\rho$ is shown in (c). The measured and modeled data are presented with the open-symbols ($\circ$) and the solid line (\full), respectively.}
  \label{fig:fig2}
\end{figure}

\citet{Kenyon} presented the dielectric data on Whitestone saturated with salty water, the system was a brine-porous-rock mixture. Here we adapted the complex permittivities of the water and the rock taken from \citet{Stroud}.  The fitting results are presented in Fig.~\ref{fig:fig2}, and the fit parameters are listed in Table~\ref{table}. It was stated that the porosity of the stone was 80\%$[=(1-q)\times100]$ in Ref. \onlinecite{Kenyon}, a value in accord with our finding $\approx0.813(=1-q\equiv1-\xi_s-\Delta\xi)$. In addition, even though the morphology of the rock was complex, the proposed expression gives evidence that some part of the brine was located in infinite pore clusters ($\xi_s>0$).

\begin{figure}[t]
  \centering
  \includegraphics[width=\sizeFIG]{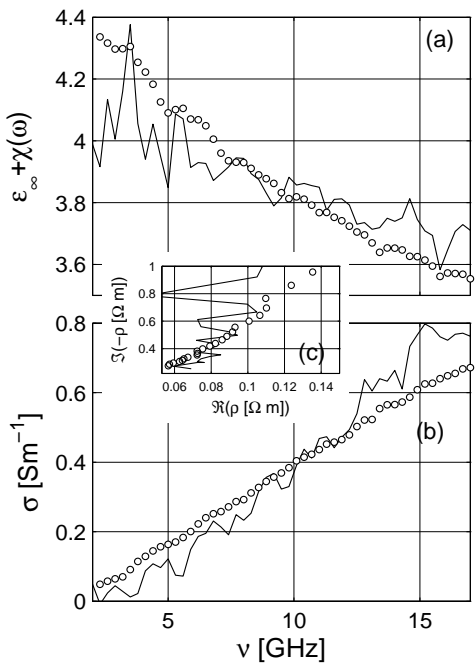}
  \caption{Frequency dependent (a) relative dielectric permittivity and (b) conductivity $\sigma$ of the glass-bead-praffin wax system. The Argand plot of the complex resistivity $\rho$ is shown in (c). The measured and modeled data are presented with the open-symbols ($\circ$) and the solid line (\full), respectively.}
  \label{fig:fig3}
\end{figure}

\citet{YoungsNano} have recently presented the dielectric properties of paraffin wax filled with metal-coated hollow glass-beads on microwave frequencies. In the present investigation, we adopt the complex permittivity of paraffin wax from the measurements, and estimate the inclusion permittivity from a dilute mixture measurement by utilizing an effective medium approximation~\cite{wiener}. Except the data at low frequencies ($\nu<5\ $GHz), Eq.~\eqref{eq:myequation2} yield results in accordance with the measurements. It is striking that the predicted concentration $q(=0.294)$ is in good agreement with the nominal volume fraction given by \citet{YoungsNano}. The fit results listed in Table~\ref{table} demostrate that most of the particles are arranged in an infinite cluster ($\xi_s>\Delta\xi$) instead of forming isolated groups. 

In this Letter, we proposed a peculiar dielectric mixture expression, which do not suffer from the applicability and inaccurate predictions of previous mixing laws in the literature  \cite{SihvolaBook}, \eg\  when inclusions  form connected network, percolation, structure, composites with arbitrary shaped inclusions and highly filled mixtures. The application of the expression to several data sets has illustrated its significance and potential in predicting and in comprehending the dielectric properties and structure of composites. We demonstrated that valuable information regarding the nature of mixtures can be obtained by the proposed expression. It can even be extended to systems exhibiting several depolarization processes,
\begin{eqnarray}
  \label{eq:general}
  \xi=\xi_s+\sum_n\Delta\xi_n[1+(\varpi x_n)^{\alpha_{n}}]^{-\beta_n},
\end{eqnarray}
which is analogous to theoretical representations of dielectric relaxation. Last but not least, it is suggested that the complex resistivity level is used while analyzing dielectric properties of composites~\cite{RossBJP}. Finally, the proposed expression can be applied to homogenization problems in other fields, \eg\ elastic modulus, thermal conductivity, magnetic permeability, \etc, where the effective property of the heterogeneous medium is of importance.

 I would like to express my thanks to Drs. Nicola Bowler (Iowa State University, USA) and Ian J. Youngs (DSTL, UK) for supplying the dielectric data on the GB/PW system and for their fruitful comments.


\end{document}